\documentclass[12pt]{article}
\usepackage{graphicx}

\begin{document}
\begin{titlepage}
\date{\today}
\title{Indirect signatures for axion(-like) particles}
\maketitle
\begin{center}
K.~Zioutas$^{1,2}$, K.~Dennerl$^3$, M.~Grande$^4$,
 D.H.H.~Hoffmann$^5$, J.~Huovelin$^6$, B.~Laki\' c$^7$, S.~Orlando$^8$,
 A.~Ortiz$^9$, Th.~Papaevangelou$^2$, Y.~Semertzidis$^{10}$, 
 Sp.~Tzamarias$^{11}$, O.~Vilhu$^6$
\vskip0.5cm
{\small \it
$^1$Department of Physics, University of Patras, Greece\\
$^2$CERN, Geneva, Switzerland\\
$^3$MPE, Garching, Germany\\
$^4$RAL, Chilton, Didcot, UK\\
$^5$IKP, TU-Darmstadt, Darmstadt, Germany\\
$^6$Observatory, University of Helsinki, Finland\\
$^7$RBI, Zagreb, Croatia\\
$^8$INAF - Osservatorio Astronomico di Palermo, Italy\\
$^9$High Altitude Observatory/N.C.A.R., Boulder, CO, USA\\
$^{10}$BNL, New York, USA\\
$^{11}$HOU, Greece\\ }

\vskip0.5cm
 E-mail :  zioutas@physics.upatras.gr
\end{center}
\newpage
\begin{abstract}
\noindent
Magnetic field dependent transient solar observations are suggestive for axion-photon oscillations with light axion(-like) particle involvement. Novel dark-moon measurements with the SMART X-ray detectors can be conclusive for radiatively decaying massive exotica like the generic solar Kaluza-Klein (KK) axions. Furthermore, the predicted intrinsic strong solar magnetic fields could be the reason of enhanced low energy axion production.
Such an axion component could be the as yet unknown origin of the strong quiet Sun X-ray luminosity at energies below $\sim 1$ keV. Solar axion telescopes should lower their threshold, aiming to copy processes that might occur near the solar surface, be it due to spontaneous or magnetically induced radiative decay of axion(-like) particles. This is motivated also by the recent claim of an axion-like particle detection by the laser experiment PVLAS.
\end{abstract}
 
\end{titlepage}
 


\section{Introduction}
The ongoing direct searches for dark matter as well as for solar 
axions are suggestive to ask whether the same detection principles are at work behind certain astrophysical observations, whose origin remains enigmatic. Nature might have implemented already a kind of fine tuning, which we may or may not be able to copy on Earth. This kind of indirect signatures for dark matter constituents might be strong at the end,
if, for example, the same exotic process can explain, in a consistent way, more celestial phenomena. This work refers to astrophysical observations of unknown origin, which can be explained with the involvement of axions or other particles with similar properties, which we call generically as axion(-like).

\section{Related astrophysical observations}
\subsection{Solar Corona  problem}
We recall that stellar observations and theory on stellar evolution cannot be reconciled with stars having atmospheres that emit X-rays
\cite{acton}. 
More specifically, the mechanism that heats the solar corona to some MK remains elusive, 
even though many possible mechanisms of coronal heating have been widely discussed in
literature;
this is one of the most challenging problems in conventional astrophysics
\cite{priest}.
This is the solar corona problem, known since 1939. To put it differently, the question is what is the origin of the sudden increase of the temperature of the solar atmosphere by a factor of $\sim 200$? At first sight this implies a violation of the second law of thermodynamics, which is difficult to accept.
A similar behaviour is encountered also at the Earth's atmosphere some 50 to 100 km height. However, the solar irradiation is at the origin of the temperature and density profiles of the terrestrial atmosphere, which resembles the solar chromosphere-corona region. This similar behaviour is suggestive for some kind of solar self-irradiation (since an external source does not exist). The radiative decay of gravitationally trapped massive particles of the type  Kaluza-Klein axions, created by the Sun itself, were considered as the generic source that heats the solar atmosphere continuously, i.e. also during quiet Sun periods
\cite{science}.
However, such particles fail to explain transient solar phenomena, asking for an additional X-ray source. 

\subsection{X-rays and solar magnetic field}
There is strong observational evidence that (transient) solar X-ray emission correlates with the local magnetic field strength (B) on the solar disk. Magnetic fields of several kGauss exist in sunspots, which are places of enhanced solar activity. It is widely accepted that the magnetic field plays a crucial role in heating the solar corona, though the exact energy release mechanism is still unknown and remains a nagging unsolved problem in astrophysics. 

These magnetic field related solar X-rays make an additional component of the solar X-ray luminosity. Having in mind the working principle of an axion magnetic helioscope like CAST 
\cite{Castpaper},
it is suggestive to assume that the celebrated solar axions with an energy 
in the $\sim $1 to $\sim$ 10 keV range, streaming out of the hot solar core, are converted into X-rays, with the Primakoff effect  occurring at the local field (B). This is actually the working principle of CAST (see contribution by Th. Dafni at these proceedings), which might take place more efficiently inside the extended surface solar magnetic fields. We do not address here a quantitative estimate, though we keep in mind that the axion-to-photon oscillation probability depends on the squared transverse magnetic field component ($\sim B^2$). Interestingly, some relevant observations with the Yohkoh telescope
\cite{bsquared1}
arrived to a soft X-ray intensity (energy below $\sim 4.4$ keV) dependence on the $B^2$ of the local surface magnetic field (B), which fits the axion-to-photon conversion inside B. Moreover, this exponent of 2 changes smoothly  within $\pm 10\%$ over the period of a whole solar cycle. This might be, at present, a possible indication for far reaching consequences in fundamental solar physics phenomena of unknown origin.  
A similar $B^2$-dependence of the X-ray intensity was observed also in ref.
\cite{bsquared2},
without arriving to more insight on the nature of such an apparent relationship. 

\begin{center}
\begin{figure}[h]
\includegraphics[width=12cm]{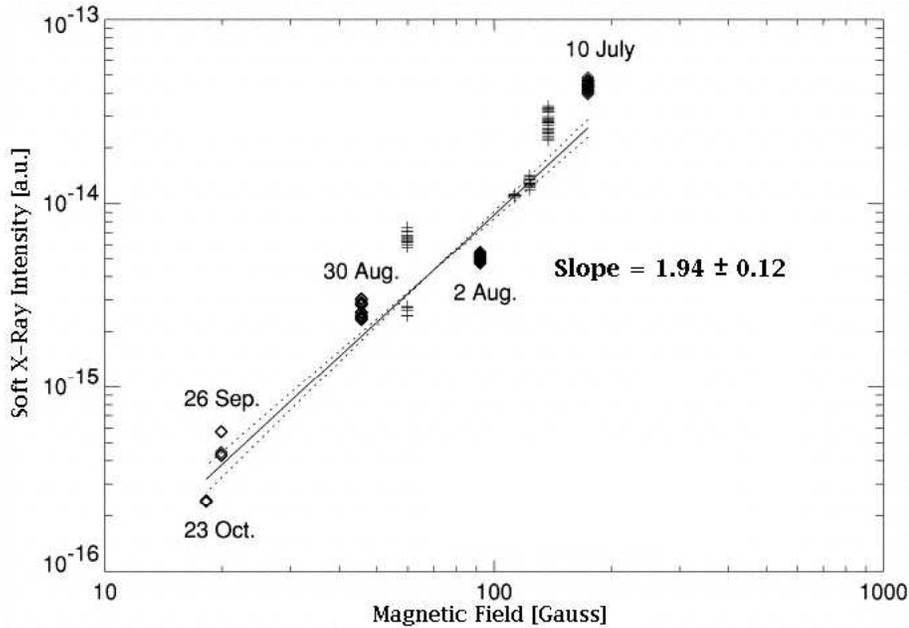}
\caption{\label{ar7978} Soft X-ray emission as a function of the local magnetic field of the Active Region AR7978 during solar minimum in 1996. The derived slope is close to 2. This figure has been taken from ref. \cite{lidia} (Permission by  Lidia van Driel-Gesztelyi).}
\end{figure}
\end{center}

The observed $L_x \sim B^2$ dependence was derived statistically, i.e. from many individual observations. Interestingly, the long-term evolution of an isolated Active Region (AR7978) during solar minimum in 1996, could be observed over few months
\cite{lidia}
confirming the above results in a more direct way. Outside flaring times, the soft X-ray luminosity from this long lived active region in loneliness,  provided (see Figure \ref{ar7978}):  
\begin{equation}
 L_x \sim B^{1.94\pm 0.12}.
\end{equation}
It is worth stressing here that in  axion helioscopes, like CAST, in the case of an axion signal, its $B^2$-dependence should be the ultimate method to be used for axion Identification. That is to say, with the unique AR7978, an efficient axion helioscope was at work for a few months cost free. 

Furthermore, what triggers the energy release of solar flares is still elusive. It is worth  further following these strong X-ray emitting events also within the axion scenario, since they correlate with the solar surface magnetic field.
Even though flares give rise to very high energy radiation, their temperature is below $\sim 20$ MK, i.e. almost equal to that of the hot core. In addition, their X-ray brightness scaled up on the whole solar surface is far below that from the inner solar nuclear engine.

\subsection{Anticorrelation of sunspot brightness with X-rays}
For sunspots, a number of fundamental questions remains unanswered, asking for an additional mechanism, which might go beyond the conventional reasoning, being  based on the magnetic field inhibited convection below those places.

Inside a magnetic field, the Primakoff effect can give rise to axion-to-photon conversion as well as to photon-to-axion backconversion as soon as photons start appearing. If the associated oscillation length is much shorter than the field length, one ends up with a mixture of  axions and photons with similar intensity each, even starting only either with axions or with photons.
This implies that part of the solar luminosity in the visible, which is streaming out from the few 100 km thick photosphere, can be temporally decreased locally, if a magnetic field with appropriate strength, inclination and length intervenes, while environmental conditions like plasma frequency, etc., can play an important role too. Each of these parameters can give rise to local/transient effects.

\begin{figure}[h]
\includegraphics[width=12cm]{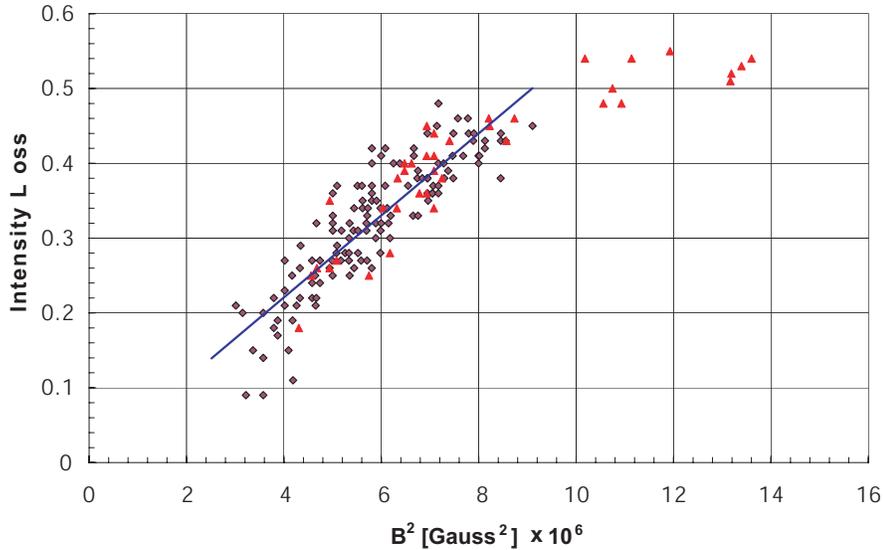}
\caption{\label{sunspots} Suppression of photon emission, in the visible, above sunspots as a function of the magnetic field squared. The solid line is a free parameter fit to the data of the underlying region.
This figure has been reconstructed from ref. \cite{solanki}.}
\end{figure}

Interestingly, the lower light intensity from sunspots decreases with the surface magnetic field squared (see Figure \ref{sunspots}) in the observed range between 1500 Gauss and 3500 Gauss by as much as 
$\sim 50 \%$ 
\cite{solanki}. 
In the axion scenario, the decreased intensity can be due to photons escaping into axions or other particles with  similar properties via the Primakoff effect occurring inside the magnetic field. 
Interestingly, as it is shown in Figure \ref{sunspots}, the magnetic field behaviour shows a $B^2$ dependence, which is characteristic for a photon-to-axion conversion. This result becomes of potential interest, when it is seen within the general dynamical behaviour of sunspots. In fact, the corona above sunspots is hotter and their photosphere underneath is cooler than near quiet sun regions, respectively
\cite{nindos}.
At first sight, the observed behaviour between two very near solar places appears contradictory. But, within the axion scenario, it is consistent, at least qualitatively: the photosphere cools because some of its photons in the visible escape as axions, while the corona gets heated by the increased X-ray emission due to enhanced axion-to-(X-ray photon) conversion inside the surface Sunspot field, as energetic axions are streaming out of the Sun.

\section{Intrinsic solar magnetic fields}
So far, the estimated solar axion production does not take into account intrinsic magnetic fields, which might reach huge strengths (see for example 
ref. \cite{intrinsic}). 
Once the local plasma frequency $\omega_{plasma}$ fits the axion rest mass, coherent photon-to-axion conversion might take place inside the magnetic field in a higher rate than that due to the incoherent Primakoff effect off the atomic Coulomb field of the hot plasma. The maximum coherence length can be equal to the photon mean free path length, i.e. a few cm inside the core to some 100 km near the surface of the Sun, provided the quasi resonance condition  $\hbar\omega_{plasma} \approx m_{axion}c^2$ applies. In particular, for axions of the KK-type, the enhanced production  due to a quasi continuous  'resonance crossing' can occur across the whole Sun. Such a process can modify the solar axion energy spectrum, depending on the field strength and its topology. Thus, a strong magnetic field outside the hot core can enhance the production of lower energy axions. The steeply increasing soft quiet Sun X-ray luminosity (of unknown origin) fits such an axion$-$magnetic-field scenario, being suggestive for further investigations.

\section{Conclusions}
The origin of the solar X-ray emission remains elusive within convetional astrophysics, with the solar corona problem insisting for $\sim 70$ years. Solar observations together with laboratory results favour novel (in)direct axion measurements in a wide energy bandwidth, utilizing various type of telescopes being Earth-bound or in space.

Thus, axion helioscopes should reach the lowest possible threshold energies.
The discussed sunspot observations along with the recent claim by the laboratory experiment PVLAS
\cite{pvlas} for an axion-like signal, motivate measurements in the visible.
At the other extreme, the search for spontaneously decaying massive exotica of the type solar KK-axions can be performed either with big analog chambers, or, with X-ray telescopes in space. We mention the dark-moon observations with the SMART observatory
\cite{smart}
: the signal depends linearly on its distance to the dark-moon, while the unknown background coming from the Moon surface is distance independent. This allows for the first time to perform conclusive background subtracted searches for massive (solar) exotica. Such a perspective demonstrates the potential importance of such measurements, and they should not be underestimated. 

The observed $B^2$-dependence of enhanced solar X-ray emission from the corona above sunspots, and, the surface brightness suppression from these as yet mysterious places, suggest that magnetic field related axion interactions could also be at work. This is then a second axion related  solar X-ray component, explaining thus transient/local phenomena, which have not been understood as yet. Note, the coupling of the generic massive solar axions of the KK-type to the magnetic field is strongly suppressed, failing to explain dynamical solar phenomema.

Observations from places beyond our Sun, like the Galactic Center, or, the Inter Cluster Medium, are of not minor mystery, since it is difficult to reconcile simple physics (at first sight) like escaping velocity, gas thermodynamics, etc., with expectation. Obviously, if we ignore, for example, a quasi electromagnetic interaction of the ubiquitous dark matter exotica out there, we end up with an unpredictable behaviour, since most astrophysical observations are based on the photon emission from those places.


We thank the CAST collaboration, since this work was motivated by the 
persisting question, how we can improve the performance of CAST.
This research was partially supported  by the  ILIAS
(Integrated Large Infrastructures for Astroparticle Science)
project  funded by the EU under contract EU-RII3-CT-2004-506222.


\end{document}